\documentclass[aps,prb,floatfix,amsmath,amssymb,preprint,eqsecnum,nofootinbib,superscriptaddress]{revtex4}
\usepackage{graphicx}
\usepackage{dcolumn}
\usepackage{bm}

\def\beq{\begin{equation}}
\def\eeq{\end{equation}}
\def\bea{\begin{eqnarray}}
\def\eea{\end{eqnarray}}

\newcommand{\br}{{\bm r}}
\newcommand{\bK}{{\bm K}}

\usepackage{setspace} 
\begin{document}

\title{DC resistivity at the onset of\\ spin density wave order in two-dimensional metals}

\author{Aavishkar A. Patel} 
\affiliation{Department of Physics, Harvard University, Cambridge MA
02138}
\author{Subir Sachdev}
\affiliation{Department of Physics, Harvard University, Cambridge MA
02138}
 \affiliation{Perimeter Institute for Theoretical Physics, Waterloo, Ontario N2L 2Y5, Canada}
\date{\today \\
\vspace{1.6in}}

\begin{abstract}
The theory for the onset of spin density wave order in a metal in two dimensions flows to strong coupling,
with strong interactions not only at the `hot spots', but on the entire Fermi surface. We advocate the computation of
 DC transport in a regime where there is rapid relaxation to local equilibrium
around the Fermi surface by processes which conserve total momentum. The DC resistivity is then controlled
by weaker perturbations which do not conserve momentum. We consider variations
in the local position of the quantum critical point, induced by long-wavelength disorder, and find a contribution to the resistivity which is linear in temperature
(up to logarithmic corrections) at low temperature. Scattering of fermions between hot spots, by short-wavelength disorder,
leads to a residual resistivity and a correction which is linear in temperature.
\end{abstract}
\maketitle

\section{Introduction}
\label{sec:intro} 

A wide variety of experiments on correlated electron compounds call for an understanding of the transport
properties of quasi-two-dimensional metals near the onset of spin density wave (SDW) order \cite{LohneysenReview,TailleferReview,pt,MatsudaReview}. Nevertheless, despite several decades of intense 
theoretical study \cite{hertz,moriya,millis,rice,rosch,AbanovChubukov,advances,ChubukovShort,kontani,max1,junhyun,max2,sdwsign,schmalian,sungsik}, the basic experimental 
phenomenology is not understood. A common feature of numerous experimental studies\cite{TailleferReview,hussey}
is a non-Fermi liquid behavior of the resistivity, which varies roughly linearly with temperature at low $T$, and more rapidly at higher $T$.

The conventional theoretical picture of transport \cite{rice,rosch} 
is that the non-Fermi liquid behavior of the electronic excitations
is limited to the vicinity of a finite number of ``hot spots'' on the Fermi surface: these are special pairs of points on the Fermi surface which are separated from each other by $\bK$, the ordering wavevector of the SDW.
The remaining Fermi surface is expected to be `cold', with sharp electron-like quasiparticles, 
and these cold quasiparticles short-circuit the electrical transport, leading to Fermi liquid behavior in the DC resistivity.

Recent theoretical works \cite{vicari,max2,schmalian} 
have called aspects of this picture into question, and argued that the cold portions
of the Fermi surface are at least `lukewarm'. Composite operators in the quantum-critical theory can lead to strong scattering
of fermionic quasiparticles at all points on the Fermi surface. Perturbatively, the deviation from Fermi liquid behavior
is strongest at the hot spots, but the quantum critical theory flows to strong coupling \cite{max1}, and so we can expect significant deviation from Fermi liquid physics all around the Fermi surface.

In the context of the DC resistivity, an important observation is that all of these deviations from Fermi liquid behavior
arise from long-wavelength processes in an effective field theory for the quantum critical point. 
Consequently, they are associated with the conservation of an appropriate momentum-like variable, and one may wonder how effective they are in relaxing the total
electrical current of the non-Fermi liquid state. For commensurate SDW with $2 \bK$ equal to a reciprocal lattice vector, it may appear that, because the interactions allow for umklapp, conservation of total momentum is not an important constraint.
However, as we will argue in more detail below, once we have re-expressed the theory in terms of the collective modes
of the effective field theory, a suitably defined momentum is conserved and its consequences have to be 
carefully tracked. It is worthwhile to note here that a similar phenomenon also appeared in the theory of transport
in the Luttinger liquid in one spatial dimension by Rosch and Andrei,\cite{roschandrei} where a single umklapp term was not sufficient to obtain a non-zero resistivity.

The present paper will address the question of the $T$ dependence of the DC resistivity at the SDW quantum critical
point using methods which represent a significant departure from the perspective of previous studies \cite{rice,rosch,advances}.
We shall employ methods similar to those used recently \cite{nematic} 
for the Ising-nematic quantum critical point, which was inspired by
analyses of transport in holographic models of metallic states \cite{hkms,hh,hartnollhofman,vegh,davison,blaketong1,blaketong2,DSZ,koushik,raghu,gouteraux,jerome,lucas},
and by Boltzmann equation studies \cite{MYC,PYM}.
Related methods have also been used for transport in non-Fermi liquids in one spatial dimension.
\cite{giamarchi,roschandrei,jung,garg}

The central assumption underlying these approaches is that the momentum-conserving interactions responsible for
the non-Fermi liquid physics are also the fastest processes leading to local thermal equilibration. We will assume here
that excitations near both the hot and lukewarm portions of the Fermi surface are susceptible to these fast processes, and are 
able to exchange momentum rapidly with each other. Then we have to look towards extraneous perturbations to relax the
total momentum, and allow for a non-zero DC resistivity. These perturbations can arise from impurities, 
from additional umklapp processes beyond those implicitly contained in the field theory, 
or from coupling to a phonon bath. Here we will focus on the impurity case exclusively, and leave the phonon contribution for future study.
The umklapp contribution can also be treated by the present methods,\cite{hartnollhofman,raghu}
and, in the approximation where cold fermions are present, yield a conventional $T^2$ resistivity.

\begin{figure}[h]
\includegraphics[width=3.3in]{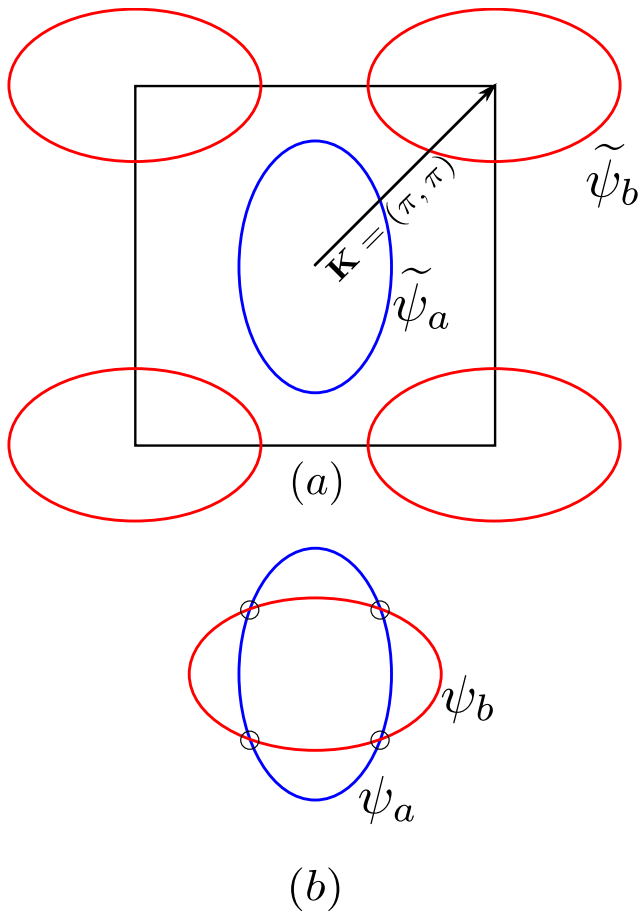}
\caption{(a) The two pockets of fermions separated by the SDW ordering wavevector $\bK=(\pi,\pi)$. (b) The resulting pair of Fermi surfaces after shifting the pocket centered at $(\pi,\pi)$ to $(0,0)$ intersect at 4 hot spots as shown.}
\label{fig:fs}
\end{figure}
For our subsequent discussion, it is useful to introduce a specific model for the SDW quantum critical point.
We find it convenient to work with a two-band model, similar to that used recently for a sign-problem-free quantum
Monte Carlo study\cite{sdwsign}. Closely related models have been used for a microscopic description of the pnictide
superconductors\cite{zlatko,eremin,vavilov,vorontsov}. As was argued in Ref.~\onlinecite{sdwsign}, we expect
our conclusions to also apply to SDW transitions in single band models because the single and two band models
have essentially the same Fermi surface structure in the vicinities of all hot spots. 
Our model begins with two species of fermions, $\widetilde{\psi}_a$, $\widetilde{\psi}_b$ 
which reside in pockets centered at $(0,0)$ and $(\pi, \pi)$ in the square
lattice Brillouin zone, as shown in Fig.~\ref{fig:fs}(a). We take the SDW ordering wavevector $\bK = (\pi, \pi)$. 
Then, we move the pocket centered at $(\pi, \pi)$ and move it to $(0,0)$ by introducing fermions
$\psi_a (\br) = \widetilde{\psi}_a (\br)$ and $\psi_b (\br) = \widetilde{\psi}_b (\br) e^{i \bK \cdot \br}$:
the Fermi surfaces for the $\psi_a$, $\psi_b$ fermions are shown in Fig.~\ref{fig:fs}(b).
The advantage of the latter representation is that the coupling of the fermions to the SDW order parameter
$\vec{\phi}$ is now local and $\br$ independent. 
So we can now write down a continuum Lagrangian for the SDW quantum critical point
in imaginary-time ($t\rightarrow -i\tau$) 
\begin{equation}
\mathcal{L}=\psi^{\dagger}\left(\partial_{\tau}-\mu_0+\left(\begin{matrix} \xi_a&0\\ 0&\xi_b \end{matrix}\right) \right)\psi + \frac{1}{2}\nabla\phi_\mu\cdot\nabla\phi_\mu+\frac{\epsilon}{2}(\partial_{\tau}\phi_\mu)(\partial_{\tau}\phi_\mu)+\frac{u}{6}\left(\phi_\mu\phi_\mu-\frac{3}{g}\right)^2+\lambda\psi^{\dagger}\phi_\mu\Gamma_\mu\psi . 
\label{LL}
\end{equation}
We have two species of spin $1/2$ fermions $(a,b)$ with chemical potential $\mu_0$ in two spatial dimensions coupled to a SO(3)  vector boson order parameter $\phi_\mu$. 
We have $\psi=\left(\begin{smallmatrix} \psi_a \\ \psi_b \end{smallmatrix}\right)$ where $\psi_{a,b}$ are two-component spinors. The matrices $\Gamma_\mu= \left(\begin{smallmatrix} 0&\sigma_\mu \\ \sigma_\mu&0 \end{smallmatrix}\right)$ with $\sigma_\mu$ as the Pauli matrices acting on the spin indices only. The dispersions of the fermions are 
\beq
\xi_{a}=-\frac{\partial_x^2}{2m_{1}}- \frac{\partial_y^2}{2m_{2}} + \ldots~,~~
\xi_{b}=-\frac{\partial_x^2}{2m_{2}}- \frac{\partial_y^2}{2m_{1}} + \ldots
\label{eq:dispersion}
\eeq
This produces two Fermi surfaces intersecting at four hot-spots (Fig.~\ref{fig:fs}(b)).
Higher-order derivatives in Eq.~(\ref{eq:dispersion}) are allowed provided additional Fermi surfaces do not appear
at larger momenta. At the critical point, we choose the value of $g$ so that the coefficient of $\phi_\mu\phi_\mu$ vanishes. We can now take the lower energy theory in the vicinities of the 4 hot spots in Fig.~\ref{fig:fs}(b),
and obtain a model identical to that studied in numerous earlier works\cite{AbanovChubukov,advances,ChubukovShort,max1,max2,sungsik}. In particular, all of the computations on the optical conductivity in Ref.~\onlinecite{max2} apply essentially unchanged to the present continuum model $\mathcal{L}$.

Now a key observation is that the resistivity of the model $\mathcal{L}$ is identically zero, $\rho (T) = 0$, at all $T$.
This follows immediately from the translational invariance of $\mathcal{L}$ and the existence of an exactly conserved
momentum which we will specify explicitly in Section~\ref{sec:sym}. 
So we must include additional perturbations to $\mathcal{L}$ will break
the continuous translational symmetry to obtain a non-zero resistivity. 
One such perturbation is a random potential, which can scatter fermions at all momenta (including $a\rightarrow b$ processes that actually change momenta by $\bK$). 
It is given by
\begin{equation}
\mathcal{L}_V=V_1(\vec{r})\psi^{\dagger}(\vec{r})\psi(\vec{r})+V_2(\vec{r})\psi^{\dagger}(\vec{r})\Gamma_0\psi(\vec{r}),
\end{equation}
where $\Gamma_0=\left(\begin{smallmatrix} 0&1\\ 1&0 \end{smallmatrix}\right)$.
The other is a random-mass term for the bosonic field:
\begin{equation}
\mathcal{L}_m=m(\vec{r})\phi_\mu(\vec{r})\phi_\mu(\vec{r}),
\end{equation}
which corresponds to a local random shift in the position of the SDW quantum critical point.
The random terms are chosen to satisfy the following upon averaging over all realizations:
\begin{eqnarray}
\langle\langle V_{1,2}(\vec{r})\rangle\rangle=0~~&;&~~\langle\langle V_{1,2}(\vec{r})V_{1,2}(\vec{r}^{\prime})\rangle\rangle=V_0^2\delta^2(\vec{r}-\vec{r}^{\prime}), \nonumber \\
\langle\langle m(\vec{r})\rangle\rangle=0~~&;&~~\langle\langle m(\vec{r})m(\vec{r}^{\prime})\rangle\rangle=m_0^2\delta^2(\vec{r}-\vec{r}^{\prime}).
\end{eqnarray}
The random-mass is expected to be a relevant perturbation to the SDW quantum critical point of $\mathcal{L}$,
and we will see that it also has a strong influence on the DC transport.

One of our main results is the following low $T$ contribution of the random-mass perturbation to the resistivity, in general accord with the scaling arguments in 
Refs.~\onlinecite{nematic} and~\onlinecite{lucas}:
\begin{equation}
\rho_m (T) \sim m_0^2 \, T^{2 (1 + \Delta -z)/z}, \label{scale1}
\end{equation}
where $z$ is the dynamic scaling exponent, and $\Delta$ is identified here with the dimension of the $\vec{\phi}^2$ operator.
In general, the latter is related to the correlation length exponent, $\nu$, via
\beq
\Delta = d+z - \frac{1}{\nu}. \label{scale2}
\eeq
Note that this contribution arises from the disorder coupling to the bosonic critical modes of the quantum critical theory,
and so is driven primarily by {\em long\/}-wavelength disorder.
In the conventional Hertz-like limit of the SDW critical point \cite{hertz,millis} we have $d=2$, $z=2$, and $\nu = 1/2$, in which case
Eqs.~(\ref{scale1},\ref{scale2}) yield $\rho_m (T) \sim T$, one of our main results. Our explicit computation also finds logarithmic corrections.
At higher temperature, we can envisage a crossover from the $z=2$ Hertz regime, to a $z=1$ Wilson-Fisher regime
\cite{sokol,georges,kachru1,kachru2,nematic,allais}: here for $d=2$, $z=1$, and\cite{holm} $\nu \approx 0.70$, Eqs.~(\ref{scale1},\ref{scale2}) yield $\rho (T) \sim T^{3.14}$. We note that a different discussion of the influence of disorder on the bosonic modes appeared recently. \cite{efetov}

We also compute the contribution of the random potential terms in $\mathcal{L}_V$ to the resistivity.
Here the dominant contribution is from the scattering of fermions between hot spots, and so this requires disorder at the {\em short\/}-wavelengths corresponding to the 
separation between the hot spots.
These lead, as expected, to a leading term which is a constant as $T \rightarrow 0$. However, we find that the leading vertex correction
has an additional contribution from scattering of fermions between hot spots which varies linearly with $T$ (up to logarithmic corrections) at low $T$. 
So we have
\beq 
\rho_V (T) \sim V_0^2 (1 + c \, T),
\eeq
for some constant $c$. 
Interestingly, we find that the vertex correction contribution is linear in $T$ even in the $z=1$ regime. 

A notable point above is that the residual resistivity arises solely from the fermionic contribution associated with $\mathcal{L}_V$, and requires short-wavelength disorder.
In contrast, the linear resistivity of $\rho_m (T)$ arises from the bosonic order parameter fluctuations coupling to long-wavelength disorder.
Thus there is no direct correlation between the magnitudes of the residual resistivity and the co-efficient of the linear resistivity.

All of the considerations of this paper also apply to other density wave transitions in two-dimensional metals, including the onset of charge density wave order.
We only require that the order parameter have a non-zero wavevector which connects two generic points on the Fermi surface, and assume that the quantum
critical theory is strongly coupled. No other feature of the spin density wave order is used in our analysis, and we focus on it mainly due to its
experimental importance. 

The body of this paper describes our computation of the DC resistivity of $\mathcal{L} + \mathcal{L}_V+\mathcal{L}_m$.
The outline is as follows: In Section~\ref{sec:sym} we discuss the continuous symmetries and derive the conserved currents of our model. In Section~\ref{sec:mem} we discuss the application of the memory matrix formalism to the calculation of the DC resistivity. In Section~\ref{sec:contrib} we obtain the contributions of the random mass term and random potential terms to the DC resistivity using the memory matrix formalism. We present details of the computations of all required quantities in the appendices.

\section{Symmetries and Noether currents}
\label{sec:sym}

The Lagrangian $\mathcal{L}$ is invariant under the following symmetries (translation, global $U(1)$ symmetry and global $SU(2)$ spin rotation symmetry):
\begin{eqnarray}
&&\vec{x}\rightarrow\vec{x}+\vec{a}, \tau\rightarrow\tau+a_0, \nonumber \\
&&\psi\rightarrow e^{i\alpha}\psi, \nonumber \\
&&\psi\rightarrow e^{\frac{i}{2}\theta_j\sigma_j}\psi, \phi_\mu\rightarrow \left(e^{i\theta_js_j}\right)_\nu\phi_\nu.
\end{eqnarray}
where $s_j$ are the generators of $SO(3)$. 

The above mentioned symmetries produce various conserved currents which may be derived using the standard Noether procedure; Translational symmetry produces
\begin{equation}
T_{ab}=\sum_{n}\left(\frac{\partial\mathcal{L}}{\partial(\partial_{a}\zeta_{n})}\partial_{b}\zeta_{n}-\partial_{a}\frac{\partial\mathcal{L}}{\partial(\partial_{c}^2\zeta_n)}\partial_{b}\zeta_n\right)-\delta_{ab}\mathcal{L},
\end{equation}
where $a,b$ are spatial indices and $\zeta_n$ are all the fields involved (in this case $\psi$ and $\phi_\mu$). Time translational invariance giving the Hamiltonian density $\mathcal{H}$ ($T_{00}$) and momentum density $\vec{\mathcal{P}}$ ($T_{0i}$) (with $\pi_\mu=-i\partial{\mathcal{L}}/(\partial(\partial_{\tau}\phi_\mu)) = - i\epsilon\partial_{\tau}\phi_\mu$, and the equal time commutation relation $[\phi_{\mu}({\vec{x}}),\pi_{\nu}(\vec{y})]=\delta^2(\vec{x}-\vec{y})\delta_{\mu\nu}$):
\begin{eqnarray}
&&\mathcal{H}(\psi,\phi_\mu,\pi_\mu)=-\psi^{\dagger}\partial_{\tau}\psi-\epsilon(\partial_{\tau}\phi_\mu)(\partial_{\tau}\phi_\mu)+\mathcal{L}(\psi,\partial_{\tau}\psi,\phi_\mu,\partial_{\tau}\phi_\mu), \nonumber \\
&&\vec{\mathcal{P}}=-\frac{i}{2}(\psi^{\dagger}\nabla\psi-\nabla\psi^{\dagger}\psi)+\pi_\mu\nabla\phi_\mu.
\end{eqnarray}
Since $\partial_{\mu}T_{\mu\nu}=0$, $\partial_{\tau}\int d^2x(\mathcal{H},\mathcal{P})=\mathrm{boundary~terms}=0$. \\

The $U(1)$ symmetry produces
\begin{equation}
j_{\mu}=\sum_{n}\left(\frac{\partial\mathcal{L}}{\partial(\partial_{\mu}\zeta_n)}-\partial_{\mu}\frac{\partial\mathcal{L}}{\partial(\partial_{\mu}^2\zeta_n)}\right)\frac{\delta\zeta_n}{\delta\alpha},
\end{equation}
which gives the current density $\vec{\mathcal{J}}$:
\begin{eqnarray}
&&\mathcal{J}_x=\frac{i}{4}\left(\frac{1}{m_1}(\partial_x\psi^{\dagger}_a\psi_a-\psi^{\dagger}_a\partial_x\psi_a)+\frac{1}{m_2}(\partial_x\psi^{\dagger}_b\psi_b-\psi^{\dagger}_b\partial_x\psi_b)\right), \nonumber \\
&&\mathcal{J}_y=\frac{i}{4}\left(\frac{1}{m_2}(\partial_x\psi^{\dagger}_a\psi_a-\psi^{\dagger}_a\partial_x\psi_a)+\frac{1}{m_1}(\partial_x\psi^{\dagger}_b\psi_b-\psi^{\dagger}_b\partial_x\psi_b)\right).
\end{eqnarray}

The $SU(2)$ symmetry produces spin currents but they cannot be used with the memory matrix approach, as explained below.

\section{Memory Matrix Approach}
\label{sec:mem}

The above theory does not possess well defined quasiparticles in two dimensions near the quantum critical point due to the strong (non-irrelevant) coupling $\lambda$, and hence it is not possible to correctly calculate transport properties like resistivity using traditional methods, as these involve doing perturbation theory in the coupling. However, the presence of a conserved total momentum $\vec{P}=\int d^2x\vec{\mathcal{P}}$, which will slowly relax if perturbations such as a weak disordered potential are applied, allows certain transport properties such as the DC resistivity to be correctly calculated using the memory matrix formalism~\cite{nematic,hartnollhofman,Forster}.

In this formalism, the conductivity tensor $\sigma_{ij}$ may be expressed as~\cite{hartnollhofman,Forster}
\begin{equation}
\sigma_{ij}(\omega)=\left(J_i\left|\frac{i}{\omega-L}\right|J_j\right),
\end{equation}
with $\vec{J}=\int d^2x\vec{\mathcal{J}}$, the Liouville super operator $L$ acting as $A(t)=e^{iHt}Ae^{-iHt}=e^{iLt}A(0)$, and the inner product of operators $(A|B)=\int_{0}^{\beta}d\tau\langle A^{\dagger}(\tau)B(0)\rangle$, with $\langle.~,~.\rangle$ denoting the connected correlation function. If the operators $A$ and $B$ have the same signature under time reversal, and the Hamiltonian is invariant under time reversal, it is easy to see that $(\dot{A}|B)=0$. Hence $(\dot{P_i}|{P_j})=0$, which simplifies the memory matrix. The dominant contributions to $\sigma(\omega)$ come from the slowly relaxing modes, which are $P_{x,y}$. Using the invariance of the Hamiltonian under $(x,y)\rightarrow(-x,y)$, the expression for the DC diagonal conductivity reduces to, to leading order in the perturbing Hamlitonian~\cite{hartnollhofman,Forster}:
\begin{equation}
\sigma_{xx}=\lim_{\omega\rightarrow0}|(J_{x}|P_{x})|^2\left(\dot{P}_{x}\left|\frac{i}{\omega-L_0}\right|\dot{P}_{x}\right)_0^{-1},
\end{equation}
where the subscript $_0$ denotes evaluation with respect to the unperturbed Hamiltonian. We then have
\begin{eqnarray}
&&\chi_{JP}=(J_x|P_x)=\int_{0}^{\beta}d\tau\langle J_x(\tau)P_x(0)\rangle, \nonumber \\
&&\sigma_{xx}=\lim_{\omega\rightarrow 0}\frac{|\chi_{JP}|^2}{\frac{1}{\omega}\int_{0}^{\infty}dt e^{i\omega t}[\dot{P}_x(t),\dot{P}_x(0)]}, \nonumber \\
&&\rho_{xx}=\mathrm{Re}\left[\frac{1}{\sigma}\right]=\lim_{\omega\rightarrow 0}\frac{\mathrm{Im}[G^{R}_{\dot{P}_x\dot{P}_x}(\omega)]}{\omega|\chi_{JP}|^2}. 
\end{eqnarray}

We compute the $\chi_{JP}$ susceptibility for $\mathcal{L}$ in Appendix~\ref{app:susc}. There we find that although the continuum limit hot-spot theory with linearized fermion dispersion has $\chi_{JP}=0$, upon including Fermi surface curvature we have $\chi_{JP} \neq 0$, even at $T=0$. We will henceforth assume that $\chi_{JP}$ is a $T$-independent non-zero constant. However, if $\chi_{JP}$ is small, then the DC resistivity $\rho (T) \sim \chi_{JP}^{-2}$ will be large, and 
 there will be a crossover to a higher $T$ regime where we
have to consider the physics of a system with $\chi_{JP}=0$: note that it is possible for such a system to have  
a non-zero resistivity even in the absence perturbations which relax momentum. 
Important, previously studied examples of theories with $\chi_{JP}=0$ are conformal field theories \cite{damle,will1,nikolai,will2} 
and it would be interesting to extend such
studies to the quantum-critical spin density wave theory \cite{max1,sungsik}.

We also see that $\chi_{SP}=0$ for the spin current due to the spin rotation symmetry of the model.

\section{Contributions to the DC Resistivity}
\label{sec:contrib}

In this section, we compute the contributions to the DC resistivity $\rho_{xx}(T)$ coming from the random-mass term and from the scattering of hot spot fermions by the random potential.  To apply the memory matrix formalism, we compute the time dependence of the conserved momentum arising from the perturbations in $\mathcal{L}_V+\mathcal{L}_m$. Using $\dot{P}_x=i[H,P_x]$, we obtain 
\begin{equation}
\dot{P}_x=-i \int \frac{d^2q~d^2k}{(2\pi)^4} k_x \left[V_1(\vec{k})\psi^{\dagger}(\vec{q}+\vec{k})\psi(q)+V_2(\vec{k})\psi^{\dagger}(\vec{q}+\vec{k})\Gamma_0\psi(q)+m(\vec{k})\phi_\mu(\vec{q})\phi_\mu(-\vec{q}-\vec{k})\right],
\end{equation}
giving
\begin{equation}
\mathrm{Im}[G^R_{\dot{P}_x,\dot{P}_x}(\omega)]=\mathrm{Im}\left[\int\frac{d^2k}{(2\pi)^2}k_x^2\left(V_0^2(\Xi_1^R(\vec{k},\omega)+\Xi_2^R(\vec{k},\omega))+m_0^2\Pi^R(\vec{k},\omega)\right)\right],
\label{eq:gpxdot}
\end{equation}
where $\Xi_{1,2}^R$ are the retarded Green's functions for $\psi^{\dagger}\psi$ and $\psi^{\dagger}\Gamma_0\psi$ respectively and $\Pi^R$ is the retarded Green's function for $\phi_\mu\phi_\mu$.

\subsection{Random-Mass Term}

\begin{figure}[h]
\includegraphics[width=6.5in]{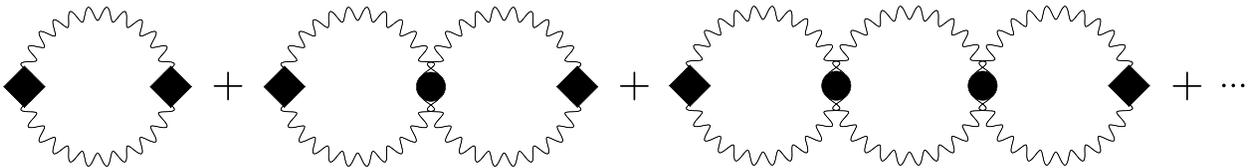}
\caption{Resummation of graphs to obtain the Green's function for $\phi_\mu\phi_\mu$. The diamonds denote $\phi_\mu\phi_\mu$ operators and the circles denote the quartic interaction. The wavy lines represent the vector boson propagators.}
\label{fig:phi2graphs}
\end{figure}

We use the following form for the vector boson propagator, which is derived in Appendix~\ref{app:rt}:
\begin{equation}
D_{\mu\nu}(\vec{q},i \omega_{q})=\frac{\delta_{\mu\nu}}{q^2+\epsilon\omega_q^2+\gamma|\omega_q|+R(T)},
\label{eq:bpr}
\end{equation}
where $R(T)$ is a positive-definite mass term at finite temperature which is computed in  Appendix~\ref{app:rt}.
The Green's function for $\phi_\mu\phi_\mu$ may be obtained by resumming the graphs shown in Fig.~\ref{fig:phi2graphs}; these are precisely the graphs
that have to be summed at leading order in a large $N$ expansion in which $\phi_\mu$ has $N$ components. We obtain
\begin{equation}
\Pi(\vec{k},i\Omega)=\frac{2\tilde{\Pi}(\vec{k},i\Omega)}{1-(20/3)u\tilde{\Pi}(\vec{k},i\Omega)},
\end{equation}
where
\begin{equation}
\tilde{\Pi}(\vec{k},i\Omega)=T\sum_{\omega_q}\int\frac{d^2q}{(2\pi)^2}D_{\mu\nu}(\vec{q},i\omega_q)D_{\nu\mu}(\vec{q}+\vec{k},i\omega_q+i\Omega).
\end{equation}

Then we have, for large $u$,
\begin{equation}
\lim_{\omega\rightarrow0}\frac{1}{\omega}\mathrm{Im}[\Pi^R(\vec{k},\omega)]=\lim_{\omega\rightarrow0}\frac{9}{200u^2\omega}\frac{\mathrm{Im}[\tilde{\Pi}^R(\vec{k},\omega)]}{\mathrm{Re}[\tilde{\Pi}^R(\vec{k},\omega)]^2},
\end{equation}

The $z=2$ regime may be accessed by sending $\epsilon\rightarrow0$ with $\gamma\neq0$. Then we have (See Appendix~\ref{app:random} for computations)
\begin{equation}
\rho_{xx}(T)=\lim_{\omega\rightarrow0}\frac{m_0^2}{\omega|\chi_{JP}|^2}\int^\Lambda\frac{d^2k}{(2\pi)^2}k_x^2\mathrm{Im}[\Pi^R(\vec{k},\omega)]\approx\frac{m_0^2\gamma^3 T}{u^2|\chi_{JP}|^2}\left[c_1+ c_2 \ln\left(\frac{\Lambda^2}{\gamma T}\right)\right],
\end{equation}
where $\Lambda$ is a momentum cutoff that is much larger than any other scale in the problem, and $c_1$, $c_2$ have only very slow log-log dependences on $T$. 

In the $\gamma\rightarrow0$ limit with $\epsilon\neq0$, $z=1$. In this regime, all the momentum integrals involved converge (See Appendix~\ref{app:random}). 
We get
\begin{eqnarray}
&&\lim_{\omega\rightarrow0}\frac{\mathrm{Im}[\tilde{\Pi}^R(\vec{k},\omega)]}{\omega}=\frac{1}{\epsilon T^2}F\left(\frac{k^2}{\epsilon T^2}\right), \nonumber \\
&&\lim_{\omega\rightarrow0}\mathrm{Re}[\tilde{\Pi}^R(\vec{k},\omega)]=\frac{1}{\epsilon T}G\left(\frac{k^2}{\epsilon T^2}\right). 
\end{eqnarray}
Thus we can cast the integral for $\rho_{xx}(T)$ in terms of a dimensionless momentum $\vec{k}^{\prime}$ and obtain
\begin{equation}
\rho_{xx}(T)=\frac{9m_0^2\epsilon^3T^4}{200u^2|\chi_{JP}|^2}\int\frac{d^2k^{\prime}}{(2\pi)^2}k^{\prime 2}_x\frac{F(k^{\prime 2})}{G(k^{\prime 2})^2}=2.42\frac{m_0^2\epsilon^3T^4}{u^2|\chi_{JP}|^2}.
\end{equation}

\begin{figure}[h]
\includegraphics[width=3.5in]{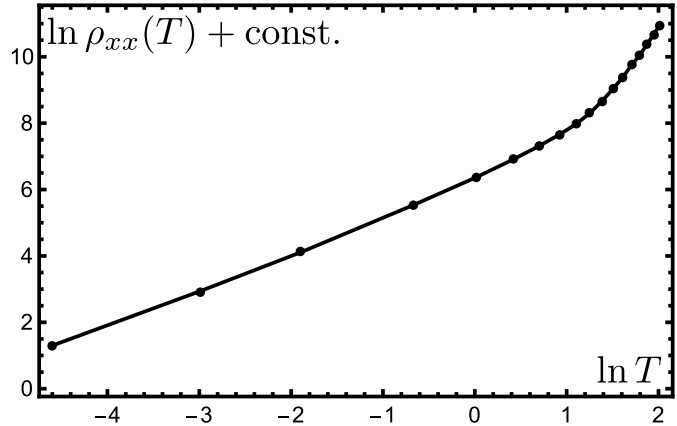}
\caption{Temperature driven crossover in the scaling of the random-mass contribution to $\rho_{xx}(T)$ from $T$ to $T^4$ as $T$ is increased. Here, $\gamma=1$, $\epsilon=1$ and the momentum cutoff $\Lambda=100$.}
\label{fig:crossover}
\end{figure}

We also obtain a temperature driven crossover in the scaling of $\rho_{xx}(T)$ when both $\epsilon\neq0$ and $\gamma\neq0$. We have $\rho(T)\sim T$ in the $z=2$ regime at low $T$ and $\rho(T)\sim T^4$ in the $z=1$ regime at high $T$, as shown in Fig.~\ref{fig:crossover}. The $T^4$ behavior agrees with Eqs.~(\ref{scale1},\ref{scale2}) with the large $N$ value of the exponent $\nu=1$.

\subsection{Fermionic Contributions}
Since the boson couples most strongly to the fermions near the hot spots, we expect the most significant non Fermi liquid contributions to the resistivity to come from the scattering of these hot spot fermions by the random potential and not involve the cold fermions elsewhere on the Fermi surfaces. The random potential can scatter these fermions between hot spots, which results in a large momentum transfer, or within the same hot spot, with a much smaller momentum transfer. Since the expression for the resistivity contribution contains a factor of $k^2$, we expect the contributions due to inter hot spot scattering to be much larger than those due to intra hot spot scattering. 

\begin{figure}[h]
\includegraphics[width=6in]{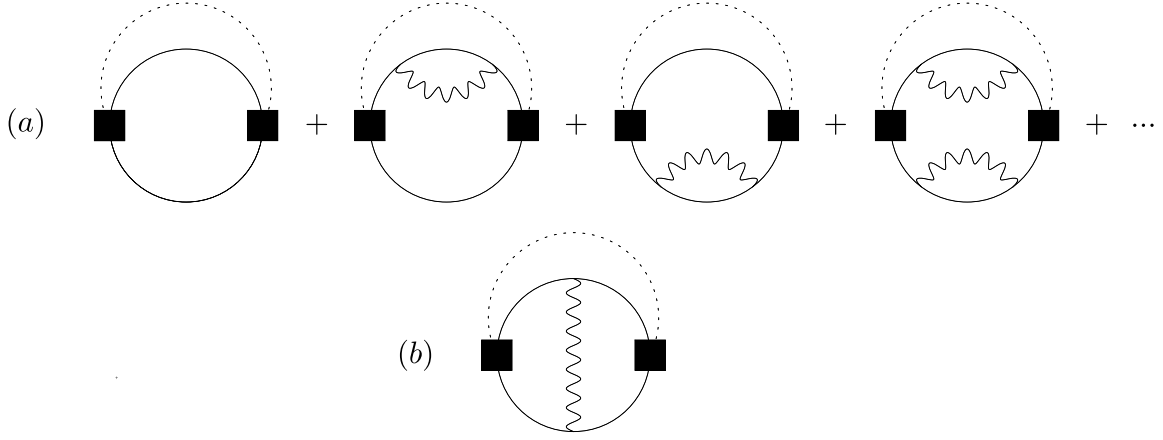}
\caption{Graphs for the contribution to $G_{\dot{P}_x,\dot{P}_x}(i\Omega)$ due to inter hot spot scattering. The vertices provide factors of $Q^{ij}V_0+Q^{ij}V_0\Gamma_0$. The solid lines are fermion propagators and the wavy lines are vector boson propagators. The dotted lines carry internal momentum and the external bosonic Matsubara frequency $i\Omega$, and have propagators equal to $1$. The first graph in the series of graphs in $(a)$ is the free fermion contribution. The subsequent graphs represent the corrections due to renormalization of the fermion propagators at one loop, but evaluate to $0$ due to factors of $\int d\xi/(i\omega-\xi)^m=0$, $m\in\mathbb{Z}$ and $m\geq2$. The graph in $(b)$ is the simplest vertex correction. Here too, for the same reason, further graphs of the same type but with self-energy rainbows on the fermion propagators also evaluate to $0$.}
\label{fig:interhsgraphs}
\end{figure}

Considering pairs of hot spots $(i,j)$, $i\neq j$ separated by vectors $\vec{Q}^{ij}$ in momentum space, we expand the momentum $\vec{k}$ transferred by the random potential about $\vec{Q}^{ij}$ in Eq.~\ref{eq:gpxdot} to obtain, to leading order, the contribution to $G_{\dot{P}_x,\dot{P}_x}$ from inter hot spot scattering
\begin{eqnarray}
&&G_{\dot{P}_x,\dot{P}_x}(i\Omega)=V_0^2\sum_{i, j, i\neq j}Q^{ij2}_x\int_0^{1/T} d\tau\Bigg[\Big\langle\psi^\dagger_j(\vec{r}=0,\tau)\psi_i(\vec{r}=0,\tau)\psi^\dagger_i(\vec{r}=0,0)\psi_j(\vec{r}=0,0)\Big\rangle+ \nonumber \\
&&\Big\langle\psi^\dagger_j(\vec{r}=0,\tau)\Gamma^0\psi_i(\vec{r}=0,\tau)\psi^\dagger_i(\vec{r}=0,0)\Gamma^0\psi_j(\vec{r}=0,0)\Big\rangle\Bigg]e^{i\Omega\tau},
\label{ihsgf}
\end{eqnarray}
where the subscripts now denote that the fermions belong to a particular hot spot, i.e. $\psi_i=\left(\begin{smallmatrix} \psi_{ia} \\ \psi_{ib} \end{smallmatrix}\right)$ and $\psi_{ia,ib}$ are two-component spinors. This leads to the graphs shown in Fig.~\ref{fig:interhsgraphs}. The fermion dispersions are now linearized about the hot spots:
\begin{equation}
\xi_{i\alpha}(\vec{k})=\vec{v}_{i\alpha}\cdot\vec{k}.
\end{equation}

The first (free fermion) graph in Fig.~\ref{fig:interhsgraphs}(a) gives
\begin{equation}
\rho_{xx}(T)=-2\pi\frac{V_0^2\Lambda^2}{|\chi_{JP}|^2}\sum_{i, j, i\neq j}Q^{ij2}_x\sum_{\alpha,\beta}\int \frac{d\xi_{i\alpha}}{(2\pi)^4v_{i\alpha}v_{i\beta}}n_F^\prime(\xi_{i\alpha})=\frac{V_0^2\Lambda_{\parallel}^2}{(2\pi)^3|\chi_{JP}|^2}\sum_{i, j, i\neq j}\sum_{\alpha,\beta}\frac{Q^{ij2}_x}{v_{i\alpha}v_{i\beta}},
\end{equation}
which is simply a temperature-independent constant. Here the indices $\alpha,\beta$ run over the two fermion types $a,b$, and $\Lambda_{\parallel}\ll Q^{ij}$ is a cutoff for the momentum components parallel to the Fermi surfaces at the hot spots. The subsequent graphs in Fig.~\ref{fig:interhsgraphs}(a) all contain factors of the form $\int d\xi/(i\omega-\xi)^m$, where $m$ is an integer $\geq2$, coming from the fermion propagators separated by self energy rainbows, and hence evaluate to zero. 
The leading vertex correction is given by the graph in Fig.~\ref{fig:interhsgraphs}(b). Again, for the same reason, we can get away with using the bare fermion propagators instead of the one loop renormalized ones. We compute this correction in Appendix~\ref{app:vertex}. In the $z=2$ limit, we obtain 
\begin{equation}
\rho_{xx}(T)\sim \mathrm{const.}+bT+c\frac{T}{\ln(\Lambda^2/(\gamma T))},
\end{equation}
which also contains terms that scale linearly in $T$. 
In the $z=1$ limit, we have
\begin{equation}
\rho_{xx}(T)\sim \mathrm{const.}+b^\prime T,
\end{equation}
which is still linear in $T$. Other corrections whose graphs contain fermion loops connected by boson propagators are less significant: due to momentum conservation at each vertex, some of these boson propagators must carry a large momentum of the order of $\vec{Q}^{ij}$, hence suppressing their contribution. Also, graphs having a single fermion loop that runs through both the external vertices, but containing multiple boson propagators which could be attached in any way, will always have the aforementioned factors that evaluate to zero once all the boson momenta and frequencies are set to zero, thus suppressing their most singular contributions.
\section{Conclusions}
\label{sec:conc}

This paper has proposed a perspective on DC transport in the vicinity of a spin-density wave quantum critical point in two dimensional metals; the results
can also apply to other density wave transitions of metals in two dimensions.
Whereas previous perturbative approaches\cite{rice,rosch} started from a quasiparticle picture which eventually breaks down at
hot spots on the Fermi surface, we have argued for a strong-coupling perspective in which no direct reference is made to quasiparticles.
Instead, we assume that strong interactions cause rapid relaxation to local thermal equilibrium, and the flow of electrical current
is determined mainly by the relaxation rate of a momentum which is conserved by the strong interactions.
We used weak disorder as the primary perturbation responsible for momentum relaxation, and then obtained a formally exact expression
for the resistivity in terms of two-point correlators of the strongly-interacting and momentum-conserving theory.

Our final results were obtained by an evaluation of such two-point correlators. Here, we used a simple large $N$ expansion, and found a resistivity
that varied linearly with $T$. Clearly, an important subject for future research is to evaluate these correlators by other methods which
are possibly more reliable in the strong-coupling limit.

Our computations also found distinct sources for the residual resistivity and the co-efficient of the linear $T$ term in the resistivity.
The residual resistivity is entirely fermionic, and arises from scattering between well-separated points on the Fermi surface, induced
by short-wavelength disorder. In contrast, the linear resistivity has a bosonic contribution from long-wavelength disorder. 
Moreover, the latter can be strongly enhanced in systems with small $\chi_{JP}$, the cross-correlator between the total momentum and the total current.

For experimental applications, BaFe$_2$(As$_{1-x}$P$_x$)$_2$ offers probably the best testing ground so far for our theory:
this material has a spin density wave quantum critical point near $x=0.3$, and a clear regime of linear-in-$T$ resistivity above it
\cite{Kasahara,Analytis}. It would be interesting to carry out these experiments while carefully reducing the degree of 
{\em long\/}-wavelength disorder, including grain boundaries and dislocations. Our theory implies that the co-efficient
of the linear-in-$T$ resistivity should decrease in such sample. Note also our argument above that the residual resistivity
cannot be used as a measure for the degree of disorder (as is often done); the residual resistivity is mainly
sensitive to {\em short\/}-wavelength disorder.

\subsection*{Acknowledgements}

We thank A.~Chubukov, S.~Hartnoll, R.~Mahajan, and A.~Rosch for useful discussions.
This research was supported by the NSF under Grant DMR-1360789, the Templeton foundation, and MURI grant W911NF-14-1-0003 from ARO.
Research at Perimeter Institute is supported by the Government of Canada through Industry Canada 
and by the Province of Ontario through the Ministry of Research and Innovation. 

\appendix

\section{Susceptibilities}
\label{app:susc}

The susceptibility $\chi_{JP}$ is taken to be the free fermion susceptibility at leading order and is thus given by
\begin{equation}
\chi_{JP}=-2\int \frac{d^2q}{(2\pi)^2}~q_x^2 \left(\frac{1}{2m_1}\frac{\partial n_F(\xi_a(\vec{q}))}{\partial \xi_a(\vec{q})}+\frac{1}{2 m_2}\frac{\partial n_F(\xi_b(\vec{q}))}{\partial \xi_b(\vec{q})}\right).
\end{equation}
defining coordinates $q_x=(2m_{1,2})^{1/2}q_{1,2}\cos\theta $,~$q_y=(2m_{2,1})^{1/2}q_{1,2}\sin\theta$,~$\theta\in[0,2\pi)$,~$q_{1,2}\in[0,\infty)$, so that $\xi_a=q_1^2$ and $\xi_b=q_2^2$ the integral can be evaluated exactly to give
\begin{equation}
\chi_{JP}=\frac{\sqrt{m_1 m_2}}{\pi} T \ln(1+e^{\frac{\mu_0}{T}})\approx \frac{\sqrt{m_1 m_2}}{\pi}\mu_0+O(T^3)+...~,
\end{equation}
where $\mu_0\gg T$ is the chemical potential for the fermions, and hence $\chi_{JP}$ is treated as a temperature-independent constant.

Both the Hamiltonian and $\vec{P}$ are invariant under $SU(2)$ spin rotation, but the spin current transforms as a vector. Hence it may be easily seen that $\chi_{SP}=0$ since the contributions from states with opposite spins will cancel. 

The linearized hot spot model has an emergent $SU(2)$ particle-hole symmetry~\cite{max1,max2}, and one obtains (with the hot spots indexed by $l$ and the fermion types indexed by $a$) 
\begin{eqnarray}
&&\vec{\mathcal{J}}=\frac{1}{2}\sum_{l,a}\vec{v}^{~l}_a\Psi^{l\dagger}_a\sigma_z \Psi^l_a, \nonumber \\
&&\vec{\mathcal{P}}=\frac{i}{4}\sum_{l,a}(\nabla\Psi^{l\dagger}_a\Psi^l_a-\Psi^{l\dagger}_a\nabla \Psi^l_a),
\end{eqnarray}
where $\Psi^l_a=\left(\begin{smallmatrix} \psi^l_a \\ i\tau_y\psi^{l\dagger}_a \end{smallmatrix}\right)$, $\psi^l_a$ are two-component spinors, the $\tau$ matrices act only on the spin indices and the $\sigma$ matrices act only on the particle hole indices. The Lagrangian is invariant under the $SU(2)$ transformations  $\Psi^l_a\rightarrow U^l \Psi^l_a = e^{i\vec{\theta}^l.\vec{\sigma}/2}\Psi^l_a$ that rotate particles into holes. One can always choose $U^l$ (for example $U^l=i\sigma_x$) such that $\vec{J}\rightarrow-\vec{J}$ and $\vec{P}\rightarrow\vec{P}$, which implies that $\chi_{JP}=0$ in this case. If a curvature of the Fermi surface is introduced (the dispersion modified to $\xi^l_a(\vec{q})=\vec{v}^l_a\cdot\vec{q}+q_x^2/(2m^l_{ax})+q_y^2/(2m^l_{ay})$), this particle-hole symmetry is broken. We then have
\begin{equation}
\chi_{JP}=-2\sum_{l,a}\left[\int^{\Lambda}_{-\Lambda}\int^{\Lambda}_{-\Lambda}\frac{dq_x dq_y}{(2\pi)^2}q_x\left(v^l_{ax}+\frac{q_x}{2m^l_{ax}}\right)n^{\prime}(\xi^l_a(\vec{q}))\right].
\end{equation}
The particle-hole symmetric regularization is chosen to make the integral vanish when the quadratic terms from the dispersion are removed, as is required by the particle-hole symmetry in that case. The integral can now be expanded in $1/m^l_{ax,y}$ to give
\begin{equation}
\chi_{JP}=\sum_{l,a}\left[\frac{C_1(\vec{v}^l_a,\Lambda,T)}{m^l_{ax}}+\frac{C_2(\vec{v}^l_a,\Lambda,T)}{m^l_{ay}}+...\right],
\end{equation} 
and hence the non-zero contributions are linear in the curvature to leading order. We emphasize here that this addition of a small curvature to the linear hot spot model is very different from the case of the two band model used throughout the paper, which has curvature built in from the beginning, and hence does not have a small value of $\chi_{JP}$ that is perturbative in the curvature.

\section{Computation of $R(T)$}
\label{app:rt}

\begin{figure}[h]
\includegraphics[width=3.5in]{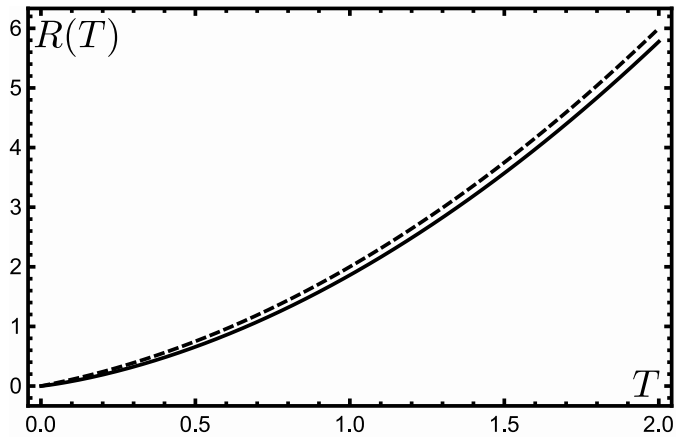}
\caption{Numerical solution (solid) of Eq.~\ref{eq:rt2}, and $\gamma T+\epsilon T^2$ (dashed), for $\epsilon=1$ and $\gamma=1$.}
\label{fig:rt}
\end{figure}

Starting with our continuum model described by Eq.~(\ref{LL}), we follow the Hertz strategy and integrate out the fermions to one loop order: As usual, only the coupling to the fermions near the hot spots modifies the boson propagator. We then consider the vector boson to have $N$ components instead of 3 for the purpose of this computation, and subsequently take a large $N$ limit. The effective Hertz action for the boson field then is
\begin{equation}
S_B=\int \frac{d^2q}{(2\pi)^2}\sum_{\omega_q}\frac{1}{2}\left[\phi_\mu(\vec{q},\omega_q)(q^2+\gamma|\omega_q|+\epsilon\omega_q^2)\phi_\mu(-\vec{q},-\omega_q)\right]+ \int d^2x~d\tau \frac{u}{2N}\left(\phi_\mu\phi_\mu-\frac{N}{g}\right)^2.
\end{equation}
Decoupling the quartic interaction using an auxiliary field $\eta$ gives
\footnotesize
\begin{equation}
S_B=\int \frac{d^2q}{(2\pi)^2}\sum_{\omega_q}\frac{1}{2}\left[\phi_\mu(\vec{q},\omega_q)(q^2+\gamma|\omega_q|+\epsilon\omega_q^2)\phi_\mu(-\vec{q},-\omega_q)\right]+ \int d^2x~d\tau\frac{1}{2}\left[\frac{i\eta}{\sqrt{N}}\left(\phi_\mu\phi_\mu-\frac{N}{g}\right)+\frac{\eta^2}{4u}\right]. 
\end{equation}
\normalsize
We now take $u\rightarrow\infty$, making the above action equivalent to that for an $O(N)$ non-linear sigma model with a fixed length constraint. Considering $\eta$ to be constant, we integrate out $\phi_\mu$ to obtain the one loop (equivalently $N=\infty$) effective potential density for $\eta$:
\begin{equation}
\mathcal{V}_{eff}=\frac{i\eta\sqrt{N}}{2g}-\frac{N}{2} \int \frac{d^2q}{(2\pi)^2}T\sum_{\omega_q}\ln\left(q^2+\gamma|\omega_q|+\epsilon\omega_q^2+\frac{i\eta}{\sqrt{N}}\right),
\end{equation}
using $i\eta/\sqrt{N}=R(T)$ and $R(0)=0$ at the critical point $g=g_c$ and minimizing this yields (while approaching the critical point from the $g>g_c$ side)
\begin{eqnarray}
&&\int \frac{d^2q}{(2\pi)^2}T\sum_{\omega_q}\frac{1}{q^2+\gamma|\omega_q|+\epsilon\omega_q^2+R(T)}=\frac{1}{g_c+0^+}, \nonumber \\
&&\int \frac{d^2q}{(2\pi)^2}T\sum_{\omega_q}\frac{1}{q^2+\gamma|\omega_q|+\epsilon\omega_q^2+R(T)}-\int \frac{d^2q}{(2\pi)^2}\int \frac{d\omega_q}{2\pi}\frac{1}{q^2+\gamma|\omega_q|+\epsilon\omega_q^2+\delta_+}=0.~~~~~~
\label{eq:rt}
\end{eqnarray}
Where $\delta_+$ is a small positive regulator. We subtract the following from the first term in the last line of the above (and add it to the second term):
\begin{equation}
\int \frac{d^2q}{(2\pi)^2}\int \frac{d\omega_q}{2\pi}\frac{1}{q^2+\gamma|\omega_q|+\epsilon\omega_q^2+R(T)};
\label{eq:rtreg}
\end{equation}
The frequency summation in the first term is carried out by analytically continuing $|\omega|$ using the following identities
\begin{equation}
|\omega|=-\frac{i\omega}{\pi}\int_{-\infty}^{\infty}\frac{dx}{x-i\omega}~~;~~ \mathrm{sgn}(\omega)=-\frac{i}{\pi}\int_{-\infty}^{\infty}\frac{dx}{x-i\omega},
\label{eq:ids}
\end{equation}
which gives
\begin{equation}
\frac{1}{q^2+\gamma|\omega_q|+\epsilon\omega_q^2+R(T)}\rightarrow\frac{1}{q^2-\epsilon z^2-i\gamma z\mathrm{sgn}(\mathrm{Im}[z])+R(T)},
\end{equation}
and avoiding the discontinuity along the real axis in the contour integration over $z$ (The function has no poles as $R(T),\gamma,\epsilon>0$). We obtain 
\begin{eqnarray}
&&T\sum_{\omega_q}\frac{1}{q^2+\gamma|\omega_q|+\epsilon\omega_q^2+R(T)}=2\int_{0}^{\infty} \frac{d\omega}{\pi} \frac{\gamma \omega}{(q^2-\epsilon\omega^2+R(T))^2+\gamma^2\omega^2}n_B(\omega)+ \nonumber \\
&&\int_{0}^{\infty} \frac{d\omega}{\pi} \frac{\gamma \omega}{(q^2-\epsilon\omega^2+R(T))^2+\gamma^2\omega^2}.
\end{eqnarray}
The limit $\delta_+\rightarrow0$ can be taken at the end without any disastrous consequences. Finally, we obtain:
\begin{eqnarray}
&&4\epsilon \int_0^\infty d\omega \left[\frac{\pi}{2}-\tan^{-1}\left(\frac{R(T)-\epsilon\omega^2}{\gamma\omega}\right)\right]n_B(\omega) + \gamma \ln \left(\frac{R(T) \epsilon}{\gamma^2} \right) \nonumber \\
&&+\sqrt{4 R(T) \epsilon -\gamma ^2} \left(2\tan ^{-1}\left(\frac{\gamma}{\sqrt{4 R(T) \epsilon -\gamma ^2}}\right)-\pi  \mathrm{sgn}\left(4 R(T) \epsilon -\gamma ^2\right)\right)=0.
\label{eq:rt2}
\end{eqnarray}
This may be solved numerically for $R(T)$, however one finds that (See Fig.~\ref{fig:rt}), to a good approximation, $R(T)$ is described by the simple form $\gamma T+ \epsilon T^2$ at intermediate values of $T$. In the $z=2$ limit ($\epsilon\rightarrow0$), we have ($\Lambda$ is a UV momentum cutoff required as a regulator in this limit)
\begin{equation}
R(T)\ln\left(\frac{\Lambda^2}{R(T)}\right)=2\gamma\int_0^\infty d\omega \tan^{-1}\left(\frac{\gamma\omega}{R(T)}\right)n_B(\omega),
\label{rz2}
\end{equation}
which gives
\begin{equation}
R(T)=\gamma Tf\left(\frac{\gamma T}{\Lambda^2}\right),
\end{equation}
where $f$ is a very slowly varying function with the property $f(0)=0$. We find 
\begin{equation}
f(x)\approx\frac{\pi W_0\left(\frac{1}{\pi}\ln\left(\frac{e^{\gamma_E-1}}{2\pi x}\right)\right)}{\ln\left(\frac{e^{\gamma_E-1}}{2\pi  x}\right)},
\label{fdef}
\end{equation}
where $W_0$ is the principal branch of the Lambert W function, and $\gamma_E$ is the Euler-Mascheroni constant. 
In the opposite limit of $z=1$ ($\gamma\rightarrow0$), we get the exact result \cite{CSY}
\begin{equation}
R(T)=\epsilon T^2\left[2\ln \left(\frac{\sqrt{5}+1}{2}\right)\right]^2.
\end{equation}

\section{Random Mass Computations}
\label{app:random}

We construct expressions for $\tilde{\Pi}(\vec{k},i\Omega)$ in terms of the spectral function for the vector boson Green's function:
\begin{equation}
A_{\mu\nu}(\vec{q},E)=\frac{-2\gamma E\delta_{\mu\nu}}{(q^2-\epsilon E^2+R(T))^2+\gamma^2E^2}.
\end{equation}
We have
\footnotesize
\begin{eqnarray}
&&\tilde{\Pi}(\vec{k},i\Omega)=T\sum_{\omega_q} \int \frac{d^2q~dE_1~dE_2}{(2\pi)^4} \frac{A_{\mu\nu}(\vec{q},E_1)}{i\omega_q-E_1}\frac{A_{\nu\mu}(\vec{q}-\vec{k},E_2) }{i\omega_q-i\Omega-E_2}, \nonumber \\
&&\lim_{\omega\rightarrow0}\frac{\mathrm{Im}[\tilde{\Pi}^R(\vec{k},\omega)]}{\omega}=6\int \frac{d^2q~dE_1}{(2\pi)^3} \frac{-\gamma^2E_1^2}{[(q^2+R(T)-\epsilon E_1^2)^2+\gamma^2E_1^2][((\vec{q}-\vec{k})^2+R(T)-\epsilon E_1^2)^2+\gamma^2 E_1^2]} n_B^{\prime}(E_1), \nonumber \\
&&\lim_{\omega\rightarrow0}\mathrm{Re}[\tilde{\Pi}^R(\vec{k},\omega)]=12\int \frac{d^2q~dE_1~dE_2}{(2\pi)^4} \frac{\gamma^2 E_1E_2} {[(q^2+R(T)-\epsilon E_1^2)^2+\gamma^2E_1^2][((\vec{q}-\vec{k})^2+R(T)-\epsilon E_2^2)^2+\gamma^2E_2^2]}\times \nonumber \\
&&\frac{n_B(E_2)-n_B(E_1)}{E_1-E_2}.
\end{eqnarray}
\normalsize
In the $z=2$ limit, performing the frequency integrals gives 
\footnotesize
\begin{eqnarray}
&&\lim_{\omega\rightarrow0}\frac{\mathrm{Im}[\tilde{\Pi}^R(\vec{k},\omega)]}{\omega}= \frac{3}{2\pi\gamma T}\int\frac{d^2q}{(2\pi)^2}\frac{1}{((\vec{q}-\vec{k})^2+R(T))^2-(q^2+R(T))^2}\Bigg[\frac{q^2+R(T)}{2\pi}\psi^\prime\left(\frac{q^2+R(T)}{2\pi\gamma T}\right)-\nonumber \\
&&\frac{(\vec{q}-\vec{k})^2+R(T)}{2\pi}\psi^\prime\left(\frac{(\vec{q}-\vec{k})^2+R(T)}{2\pi\gamma T}\right)+\pi\gamma^2T^2\left(\frac{1}{(\vec{q}-\vec{k})^2+R(T)}- \frac{1}{q^2+R(T)}\right)\Bigg], \nonumber \\
&&\lim_{\omega\rightarrow0}\mathrm{Re}[\tilde{\Pi}^R(\vec{k},\omega)]=\frac{6}{2\pi\gamma}\int^\Lambda \frac{d^2q}{(2\pi)^2}\frac{1}{q^2-(\vec{q}-\vec{k})^2}\Bigg[\psi\left(\frac{q^2+R(T)}{2\pi\gamma T}\right)-\psi\left(\frac{(\vec{q}-\vec{k})^2+R(T)}{2\pi\gamma T}\right)+\nonumber \\
&&\pi\gamma T \left(\frac{1}{q^2+R(T)}-\frac{1}{(\vec{q}-\vec{k})^2+R(T)}\right)\Bigg],
\end{eqnarray}
\normalsize
where $\psi$ here is the digamma function, and $\Lambda$ is a momentum cutoff. We obtain the following asymptotic forms in $k$ (only the dependence on $k,T,\Lambda$ and $\gamma$ is shown)
\begin{eqnarray}
&&\lim_{\omega\rightarrow0}\frac{\mathrm{Im}[\tilde{\Pi}^R(\vec{k},\omega)]}{\omega}\sim\frac{\gamma T}{k^4}\ln f\left(\frac{\gamma T}{\Lambda^2}\right),~~k^2\gg\gamma T, \nonumber \\
&&\lim_{\omega\rightarrow0}\frac{\mathrm{Im}[\tilde{\Pi}^R(\vec{k},\omega)]}{\omega}\sim\frac{1}{\gamma T}\left[f\left(\frac{\gamma T}{\Lambda^2}\right)\right]^{-2},~~k^2\ll\gamma T, \nonumber \\
&&\lim_{\omega\rightarrow0}\mathrm{Re}[\tilde{\Pi}^R(\vec{k},\omega)]\sim \frac{1}{\gamma},~~k^2\rightarrow\Lambda^2\gg\gamma T, \nonumber \\
&&\lim_{\omega\rightarrow0}\mathrm{Re}[\tilde{\Pi}^R(\vec{k},\omega)]\sim \frac{1}{\gamma}\left(b_1\left[f\left(\frac{\gamma T}{\Lambda^2}\right)\right]^{-1}+b_2\ln\left(\frac{\Lambda^2}{\gamma T}\right)\right),~~k^2\ll\gamma T,
\end{eqnarray}
where $f$ is the function correcting the linear dependence of $R(T)$ on $T$ defined in Eq.~(\ref{fdef}). We then have
\begin{equation}
\rho_{xx}(T)\propto \frac{m_0^2}{u^2|\chi_{JP}|^2} \lim_{\omega\rightarrow0}\frac{1}{\omega}\int_0^{\sqrt{\gamma T}}+\int_{\sqrt{\gamma T}}^\Lambda k^3 dk \frac{\mathrm{Im}[\tilde{\Pi}^R(k,\omega)]}{\mathrm{Re}[\tilde{\Pi}^R(k,\omega)]^2}.
\end{equation}
Substituting the small $k$ asymptotic forms in the first integral and the large $k$ ones in the second, and noting that $f(x)\sim\ln\ln(1/x)/\ln(1/x)$, we obtain the scaling form given in the main text to leading-log order, which agrees well with numerical evaluation of the integrals.

For the $z=1$ limit, we have,
\begin{eqnarray}
&&\lim_{\omega\rightarrow0}\frac{\mathrm{Im}[\tilde{\Pi}^R(\vec{k},\omega)]}{\omega}=\frac{3}{16\pi\sqrt{\epsilon}}\int_{-\pi/2}^{\pi/2} \frac{d\theta}{\cos^2\theta}\left[\frac{-n_B^\prime\left(\sqrt{k^2/(4\cos^2\theta)+R(T)}/\sqrt{\epsilon}\right)}{\sqrt{k^2/(4\cos^2\theta)+R(T)}}\right], \nonumber \\
&&\lim_{\omega\rightarrow0}\mathrm{Re}[\tilde{\Pi}^R(\vec{k},\omega)]=\frac{3}{2\sqrt{\epsilon}}\int\frac{d^2q}{(2\pi)^2}\frac{1}{q^2-(\vec{q}-\vec{k})^2}\Bigg[\frac{2n_B\left(\sqrt{(\vec{q}-\vec{k})^2+R(T)}\Big/\sqrt{\epsilon}\right)+1}{\sqrt{(\vec{q}-\vec{k})^2+R(T)}}- \nonumber \\
&&\frac{2n_B\left(\sqrt{q^2+R(T)}/\sqrt{\epsilon}\right)+1}{\sqrt{q^2+R(T)}}\Bigg]. 
\end{eqnarray}
These integrals are convergent, and we can thus scale out $\epsilon T^2$ after plugging in $R(T)$ to get the result in the main text.

In the crossover region between the $z=2$ to $z=1$ regime, we evaluate all integrals numerically and plug in the numerical solution for $R(T)$ at arbitrary $T$ to obtain Fig.~\ref{fig:crossover}. 

\section{Vertex Correction for Inter Hot-Spot Scattering}
\label{app:vertex}
We now compute the graph in Fig.~\ref{fig:interhsgraphs}(b), which is the leading vertex correction to the resistivity for inter hot-spot scattering. In the approximation of Eq.~\ref{ihsgf}, the momenta flowing through the upper and lower fermion lines in the graph are independent of each other. Since the bare fermion propagator depends only on the component of its momentum transverse to the fermi surface, and because the interaction with the boson switches the fermion type, we have (using the spectral representation for the boson Green's function): 
\footnotesize
\begin{eqnarray}
&&\rho_{xx}(T)=-\lim_{\omega\rightarrow0}\frac{6V_0^2\lambda^2}{\omega |\chi_{JP}|^2}\sum_{i,j,i\neq j}\sum_{\alpha,\beta}\frac{Q_{x}^{ij2}}{|\vec{v}_{i\alpha}\times\vec{v}_{i\bar{\alpha}}||\vec{v}_{j\beta}\times\vec{v}_{j\bar{\beta}}|}\mathrm{Im}\Bigg[\int\frac{d\xi_{i\alpha}d\xi_{i\bar{\alpha}}d\xi_{j\beta}d\xi_{j\bar{\beta}}dEd^2q}{(2\pi)^7}T^2\sum_{\omega_q,\eta}\frac{1}{i\omega_q-\xi_{i\alpha}} \times \nonumber \\
&&\frac{1}{i\omega_q+i\eta-\xi_{i\bar{\alpha}}-\vec{v}_{i\bar{\alpha}}\cdot\vec{q}}\frac{1}{i\omega_q+i\eta-i\Omega-\xi_{j\beta}-\vec{v}_{j\beta}\cdot\vec{q}}\frac{1}{i\omega_q-i\Omega-\xi_{j\bar{\beta}}}\frac{1}{i\eta-E}\frac{-2\gamma E}{(q^2+R(T))^2+\gamma^2E^2}\Bigg]_{i\Omega\rightarrow\omega+i0^+} \nonumber \\
&&=-\lim_{\omega\rightarrow0}\frac{6V_0^2\lambda^2}{\omega |\chi_{JP}|^2}\sum_{i,j,i\neq j}\sum_{\alpha,\beta}\frac{Q_{x}^{ij2}}{|\vec{v}_{i\alpha}\times\vec{v}_{i\bar{\alpha}}||\vec{v}_{j\beta}\times\vec{v}_{j\bar{\beta}}|}\mathrm{Im}\Bigg[\int\frac{d\xi_{i\alpha}d\xi_{i\bar{\alpha}}d\xi_{j\beta}d\xi_{j\bar{\beta}}dE}{(2\pi)^6}T^2\sum_{\omega_q,\eta}\frac{1}{i\omega_q-\xi_{i\alpha}} \times \nonumber \\
&&\frac{1}{i\omega_q+i\eta-\xi_{i\bar{\alpha}}}\frac{1}{i\omega_q+i\eta-i\Omega-\xi_{j\beta}}\frac{1}{i\omega_q-i\Omega-\xi_{j\bar{\beta}}}\frac{1}{i\eta-E}\left(\tan^{-1}\left(\frac{R(T)}{\gamma E}\right)-\frac{\pi}{2}\right)\Bigg]_{i\Omega\rightarrow\omega+i0^+},
\label{vertcor}
\end{eqnarray}
\normalsize
where the indices $\alpha,\beta$ run over the fermion types $a,b$, $\bar{a}=b$, $\bar{b}=a$, $\omega_q$ is a fermionic Matsubara frequency, and $\eta,\Omega$ are bosonic Matsubara frequencies. In the second step in the above, we have used the independence of the $\xi_{i\alpha}$'s to shift out the boson momenta entering the fermion propagators. One should note that here since all the fermion propagators have independent $\xi_{i\alpha}$'s, factors of $\int d\xi/(i\omega-\xi)^{m\geq2}=0$ do not appear even when the boson momentum and frequency go to zero, and the most singular contribution of the graph thus survives. This will not be the case for the higher order corrections mentioned at the end of this appendix.  After carrying out the frequency summations, We have to evaluate \par
\scriptsize
\begin{eqnarray}
&&\mathrm{Im}\int \frac{d\xi_{i\alpha}d\xi_{i\bar{\alpha}}d\xi_{j\beta}d\xi_{j\bar{\beta}}dE}{(2\pi)^6} \Bigg(-\frac{n_B\left(E\right) n_F\left(\xi_{i\alpha}\right)}{\left(\xi_{i\alpha}-\xi_{i\bar{\alpha}}+E\right) \left(-i \Omega +\xi_{i\alpha}-\xi_{i\bar{\alpha}}\right) \left(-i \Omega +\xi_{i\alpha}-\xi_{j\beta}+E\right)}-\nonumber \\\
&&\frac{n_B\left(E\right) n_F\left(\xi_{j\beta}-E\right)}{\left(\xi_{j\beta}-\xi_{j\bar{\beta}}-E\right) \left(i \Omega -\xi_{i\bar{\alpha}}+\xi_{j\beta}\right) \left(i \Omega -\xi_{i\alpha}+\xi_{j\beta}-E\right)} -\frac{n_B\left(E\right) n_F\left(\xi_{j\bar{\beta}}\right)}{\left(-\xi_{j\beta}+\xi_{j\bar{\beta}}+E\right) \left(i \Omega -\xi_{i\alpha}+\xi_{j\bar{\beta}}\right) \left(i \Omega -\xi_{i\bar{\alpha}}+\xi_{j\bar{\beta}}+E\right)}- \nonumber \\
&&\frac{n_B\left(E\right) n_F\left(\xi_{i\bar{\alpha}}-E\right)}{\left(-\xi_{i\alpha}+\xi_{i\bar{\alpha}}-E\right) \left(-i \Omega +\xi_{i\bar{\alpha}}-\xi_{j\beta}\right) \left(-i \Omega +\xi_{i\bar{\alpha}}-\xi_{j\bar{\beta}}-E\right)} +\frac{n_F\left(\xi_{i\bar{\alpha}}\right) n_F\left(\xi_{i\alpha}\right)}{\left(-\xi_{i\alpha}+\xi_{i\bar{\alpha}}-E\right) \left(-i \Omega +\xi_{i\alpha}-\xi_{j\bar{\beta}}\right) \left(-i \Omega +\xi_{i\bar{\alpha}}-\xi_{j\beta}\right)} - \nonumber \\
&&\frac{n_F\left(\xi_{i\bar{\alpha}}\right) n_F\left(\xi_{i\bar{\alpha}}-E\right)}{\left(-\xi_{i\alpha}+\xi_{i\bar{\alpha}}-E\right) \left(-i \Omega +\xi_{i\bar{\alpha}}-\xi_{j\beta}\right) \left(-i \Omega +\xi_{i\bar{\alpha}}-\xi_{j\bar{\beta}}-E\right)} -\frac{n_F\left(\xi_{j\beta}\right) n_F\left(\xi_{j\beta}-E\right)}{\left(\xi_{j\beta}-\xi_{j\bar{\beta}}-E\right) \left(i \Omega -\xi_{i\alpha}+\xi_{j\beta}-E\right) \left(i \Omega -\xi_{i\bar{\alpha}}+\xi_{j\beta}\right)}+ \nonumber \\
&&\frac{n_F\left(\xi_{j\beta}\right) n_F\left(\xi_{j\bar{\beta}}\right)}{\left(\xi_{j\beta}-\xi_{j\bar{\beta}}-E\right) \left(i \Omega -\xi_{i\alpha}+\xi_{j\bar{\beta}}\right) \left(i \Omega -\xi_{i\bar{\alpha}}+\xi_{j\beta}\right)} +\frac{n_F\left(\xi_{i\bar{\alpha}}\right) n_F\left(\xi_{j\bar{\beta}}\right)}{\left(i \Omega -\xi_{i\alpha}+\xi_{j\bar{\beta}}\right) \left(-i \Omega +\xi_{i\bar{\alpha}}-\xi_{j\bar{\beta}}-E\right) \left(-i \Omega +\xi_{i\bar{\alpha}}-\xi_{j\beta}\right)} \nonumber \\
&&+\frac{n_F\left(\xi_{j\beta}\right) n_F\left(\xi_{i\alpha}\right)}{\left(i \Omega -\xi_{i\bar{\alpha}}+\xi_{j\beta}\right) \left(i \Omega -\xi_{i\alpha}+\xi_{j\beta}-E\right) \left(-i \Omega +\xi_{i\alpha}-\xi_{j\bar{\beta}}\right)}\Bigg)\times\left(\tan^{-1}\left(\frac{R(T)}{\gamma E}\right)-\frac{\pi}{2}\right)\Bigg|_{i\Omega\rightarrow\omega+i0^+}, 
\end{eqnarray}
\normalsize
Using $1/(x+i0^{\pm})=\mp i \pi\delta(x)+\mathcal{P}/x$, the imaginary parts of the first eight terms inside the brackets in the above vanish. For the last term, relabeling dummy variables $\xi_{i\alpha}\leftrightarrow \xi_{i\bar{\alpha}}$, $\xi_{j\beta}\leftrightarrow \xi_{j\bar{\beta}}$ and $E\rightarrow-E$ simplifies the above expression to \par
\scriptsize
\begin{eqnarray}
&&\mathrm{Im}\int \frac{d\xi_{i\alpha}d\xi_{i\bar{\alpha}}d\xi_{j\beta}d\xi_{j\bar{\beta}}dE}{(2\pi)^6}\frac{2n_F\left(\xi_{i\bar{\alpha}}\right) n_F\left(\xi_{j\bar{\beta}}\right)}{\left(i \Omega -\xi_{i\alpha}+\xi_{j\bar{\beta}}\right) \left(-i \Omega +\xi_{i\bar{\alpha}}-\xi_{j\bar{\beta}}-E\right) \left(-i \Omega +\xi_{i\bar{\alpha}}-\xi_{j\beta}\right)}\tan^{-1}\left(\frac{R(T)}{\gamma E}\right)\Bigg|_{i\Omega\rightarrow\Omega+i0^+} \nonumber \\
&&=\pi^2 \int \frac{d\xi_{i\alpha}d\xi_{i\bar{\alpha}}d\xi_{j\beta}d\xi_{j\bar{\beta}}dE}{(2\pi)^5} n_F\left(\xi_{i\bar{\alpha}}\right) n_F\left(\xi_{j\bar{\beta}}\right)\delta(\omega-\xi_{i\alpha}+\xi_{j\bar{\beta}})\delta(-\omega+\xi_{i\bar{\alpha}}-\xi_{j\bar{\beta}}-E)\delta(-\omega+\xi_{i\bar{\alpha}}-\xi_{j\beta})\tan^{-1}\left(\frac{R(T)}{\gamma E}\right) \nonumber \\
&&=\frac{1}{32\pi^3} \int d\xi_{i\bar{\alpha}}d\xi_{j\bar{\beta}} n_F\left(\xi_{i\bar{\alpha}}\right) n_F\left(\xi_{j\bar{\beta}}\right)\tan^{-1}\left(\frac{R(T)}{\gamma(\xi_{i\bar{\alpha}}-\xi_{j\bar{\beta}}-\omega)}\right). 
\end{eqnarray}
\normalsize
If $\omega=0$ this evaluates to $0$ as the integrand is odd under $\xi_{i\bar{\alpha}}\leftrightarrow \xi_{j\bar{\beta}}$. Hence we have
\footnotesize
\begin{equation}
\rho_{xx}(T)=-\frac{3V_0^2\lambda^2}{16\pi^3|\chi_{JP}|^2}\sum_{i,j,i\neq j}Q_x^{ij2}\sum_{\alpha,\beta}\int_{-\tilde{\Lambda}}^{\tilde{\Lambda}} \frac{d\xi_{i\bar{\alpha}}d\xi_{j\bar{\alpha}}}{|\vec{v}_{i\alpha}\times\vec{v}_{i\bar{\alpha}}||\vec{v}_{j\beta}\times\vec{v}_{j\bar{\beta}}|} n_F\left(\xi_{i\bar{\alpha}}\right) n_F\left(\xi_{j\bar{\alpha}}\right)\frac{\gamma R(T)}{\gamma^2(\xi_{i\bar{\alpha}}-\xi_{j\bar{\alpha}})^2+R(T)^2},
\end{equation}
\normalsize
Where the cutoff $\tilde{\Lambda}\gg T$ is used to regulate the divergence of the integral as $\xi_{i\alpha},\xi_{i\bar{\alpha}}\rightarrow-\infty$. We decompose the integration into four quadrants and obtain to leading-log order in $T$: 
\small
\begin{eqnarray}
&&I_{++}=\int_{0}^{\tilde{\Lambda}}\int_{0}^{\tilde{\Lambda}}d\xi_{i\bar{\alpha}}d\xi_{j\bar{\alpha}}n_F\left(\xi_{i\bar{\alpha}}\right) n_F\left(\xi_{j\bar{\alpha}}\right)\frac{\gamma R(T)}{\gamma^2(\xi_{i\bar{\alpha}}-\xi_{j\bar{\alpha}})^2+R(T)^2}\approx T\left[a_1+a_2f\left(\frac{\gamma T}{\Lambda^2}\right)\right], \nonumber \\
&&I_{+-}=I_{-+}=\int_{0}^{\tilde{\Lambda}}\int_{-\tilde{\Lambda}}^{0}d\xi_{i\bar{\alpha}}d\xi_{j\bar{\alpha}}n_F\left(\xi_{i\bar{\alpha}}\right) n_F\left(\xi_{j\bar{\alpha}}\right)\frac{\gamma R(T)}{\gamma^2(\xi_{i\bar{\alpha}}-\xi_{j\bar{\alpha}})^2+R(T)^2}\approx a_3 T f\left(\frac{\gamma T}{\Lambda^2}\right), \nonumber \\
&&I_{--}^{\mathrm{reg}}=\int_{-\tilde{\Lambda}}^{0}\int_{-\tilde{\Lambda}}^{0}d\xi_{i\bar{\alpha}}d\xi_{j\bar{\alpha}}(n_F\left(\xi_{i\bar{\alpha}}\right) n_F\left(\xi_{j\bar{\alpha}}\right)-1)\frac{\gamma R(T)}{\gamma^2(\xi_{i\bar{\alpha}}-\xi_{j\bar{\alpha}})^2+R(T)^2}\approx T\left[a_4+a_5f\left(\frac{\gamma T}{\Lambda^2}\right)\right],  \nonumber \\
&&I_{--}^{\mathrm{div}}=\int_{-\tilde{\Lambda}}^{0}\int_{-\tilde{\Lambda}}^{0}d\xi_{i\bar{\alpha}}d\xi_{j\bar{\alpha}}\frac{\gamma R(T)}{\gamma^2(\xi_{i\bar{\alpha}}-\xi_{j\bar{\alpha}})^2+R(T)^2}\approx \pi\tilde{\Lambda}+2\frac{R(T)}{\gamma}\ln\left(\frac{R(T)}{\tilde{\gamma \Lambda}}\right),
\end{eqnarray}
\normalsize
where $a_1=\pi\ln(2/\sqrt{e})$, $a_4=-\pi\ln(2\sqrt{e})$, and $a_{2,3,5}$ have very slow log-log dependences on $T$. Here $f$ is the function correcting the linear dependence of $R(T)$ on $T$ and is defined in Eq.~\ref{fdef}, and $\Lambda$ is the cutoff used in Eq.~\ref{rz2}. We thus obtain
\begin{equation}
\rho_{xx}(T)\approx\frac{V_0^2\lambda^2}{|\chi_{JP}|^2}\sum_{i,j,i\neq j}Q_x^{ij2}\sum_{\alpha,\beta}\frac{1}{|\vec{v}_{i\alpha}\times\vec{v}_{i\bar{\alpha}}||\vec{v}_{j\beta}\times\vec{v}_{j\bar{\beta}}|}\left[-a\tilde{\Lambda}+bT+c\frac{T}{\ln(\Lambda^2/(\gamma T))}\right],
\end{equation}
at low $T$ where $a,b>0$ and $b,c$ have very slow log-log dependences on $T$.

In the $z=1$ limit, the factor of $\tan^{-1}(R(T)/(\gamma E))-\pi/2$ in Eq.~\ref{vertcor} is replaced with $-\pi\Theta(\epsilon(E^2-cT^2))$. Then, performing the same computation yields $\rho_{xx}(T)\sim -a^\prime\tilde{\Lambda}+b^\prime T$, $a^\prime,b^\prime>0$. 

We can also consider other graphs which have a fermion loop that runs through both the external vertices, and multiple internal boson propagators that intersect this loop at various points (For example, one such family of graphs would be the higher order graphs in the ``ladder series" of graphs, which contain multiple boson propagators connecting the upper and lower fermion lines instead of just one in the above vertex correction). The most singular contribution from these graphs would arise when the momenta and frequencies of all these internal boson propagators go to zero simultaneously: When this happens, such graphs will be given by expressions of the form
\footnotesize
\begin{eqnarray}
&&\sum_{i,j,i\neq j}\sum_{\alpha,\beta}\frac{Q_{x}^{ij2}}{|\vec{v}_{i\alpha}\times\vec{v}_{i\bar{\alpha}}||\vec{v}_{j\beta}\times\vec{v}_{j\bar{\beta}}|}T\sum_{\omega_q}\int d\xi_{i\alpha}d\xi_{i\bar{\alpha}}d\xi_{j\beta}d\xi_{j\bar{\beta}} \frac{1}{(i\omega_q-\xi_{i\alpha})^{t_1}} \frac{1}{(i\omega_q-\xi_{i\bar{\alpha}})^{t_2}} \frac{1}{(i\omega_q-i\Omega-\xi_{j\beta})^{t_3}}\times\nonumber \\
&&\frac{1}{(i\omega_q-i\Omega-\xi_{j\bar{\beta}})^{t_4}}\left(\frac{T}{R(T)}\right)^n,
\end{eqnarray}
\normalsize
where $n$ is the number of internal boson propagators. It is guaranteed that at least one of the $t$'s is $\geq2$, because at least one of the fermion lines will have more than one intersection with an internal boson propagator if there is more than one internal boson propagator. Hence this expression to evaluates to zero, and the most singular contribution vanishes.

\end{document}